%% file: main.tex
\documentclass[10pt]{book}
\usepackage{pgstylestandalone}

\title{Political Geometry}
\author{  }
\date{ }

\begin{document}
\Urlmuskip=0mu plus 1mu\relax 


 \input{guth-nieh-weighill.tex}    
 
\end{document}

%% file: guth-nieh-weighill.tex
\chapter{Three Applications of Entropy to Gerrymandering}
\label{GuthNiehWeighill}
\chapterauthor{LARRY GUTH \\ ARI NIEH \\ THOMAS WEIGHILL}
\authorheader{GUTH, NIEH \& WEIGHILL}
\titleheader{Three Applications of Entropy}

\chaptersummary{This preprint is an exploration in how a single mathematical idea-- entropy-- can be applied to redistricting in a number of ways. It's meant to be read not so much as a call to action for entropy, but as a case study illustrating one of the many ways math can inform our thinking on redistricting problems. This preprint was prepared as a chapter in the forthcoming edited volume Political Geometry, an interdisciplinary collection of essays on redistricting. (\url{mggg.org/gerrybook})}

\section{Introduction}
Redistricting primarily concerns data, typically data about people -- which political jurisdictions they reside in, what demographic groups they belong to and how they vote, to name just a few. More than raw data, though, studying redistricting is about investigating the relationships between different kinds of data: for example, how is the demographic and vote data distributed amongst the districts in a state? This is not unique to redistricting -- many scientific fields now find that their biggest questions revolve around studying relationships between large data sets. Mathematical tools originally designed for completely different purposes can help us address these questions. In this chapter, we examine a tool of this kind: information theory, and more specifically, the notion of entropy.

Mathematician Claude Shannon originally developed information theory during the Second World War to find an upper bound on how much information could be transmitted through a communication channel. One can easily see why such a bound would be important for someone working on the secret communication system used by Churchill and Roosevelt. The basic idea is that information is related to uncertainty: the fewer patterns exist in the data, the harder it is to efficiently communicate that data -- that is, there is more information we have to get across. The story goes that Shannon was persuaded by Von Neumann to call this quantity neither ``information'' nor ``uncertainty'' but rather \emph{entropy}, inspired by the thermodynamic quantity of the same name. 

Consider the following very simple example. Suppose that you want to know the party affiliation of every person in a town hall meeting. (For the sake of simplicity, pretend that only Republicans and Democrats are in attendance.) If you know beforehand that the vast majority of the attendees will be Republicans, then you can ask the Democrats to raise their hands and assume that those that didn't were Republican. Here, the effort to communicate the data is measured in how many people had to raise their hands. In this case, not much effort was required on average by the audience -- the entropy in the party-affiliation data is low. If, on the other hand, the room was a fifty-fifty split, then more hands would have to be raised, resulting in a greater effort per person -- in this case, the party-affiliation data has high entropy.

\begin{center}
    \includegraphics[width=\textwidth]{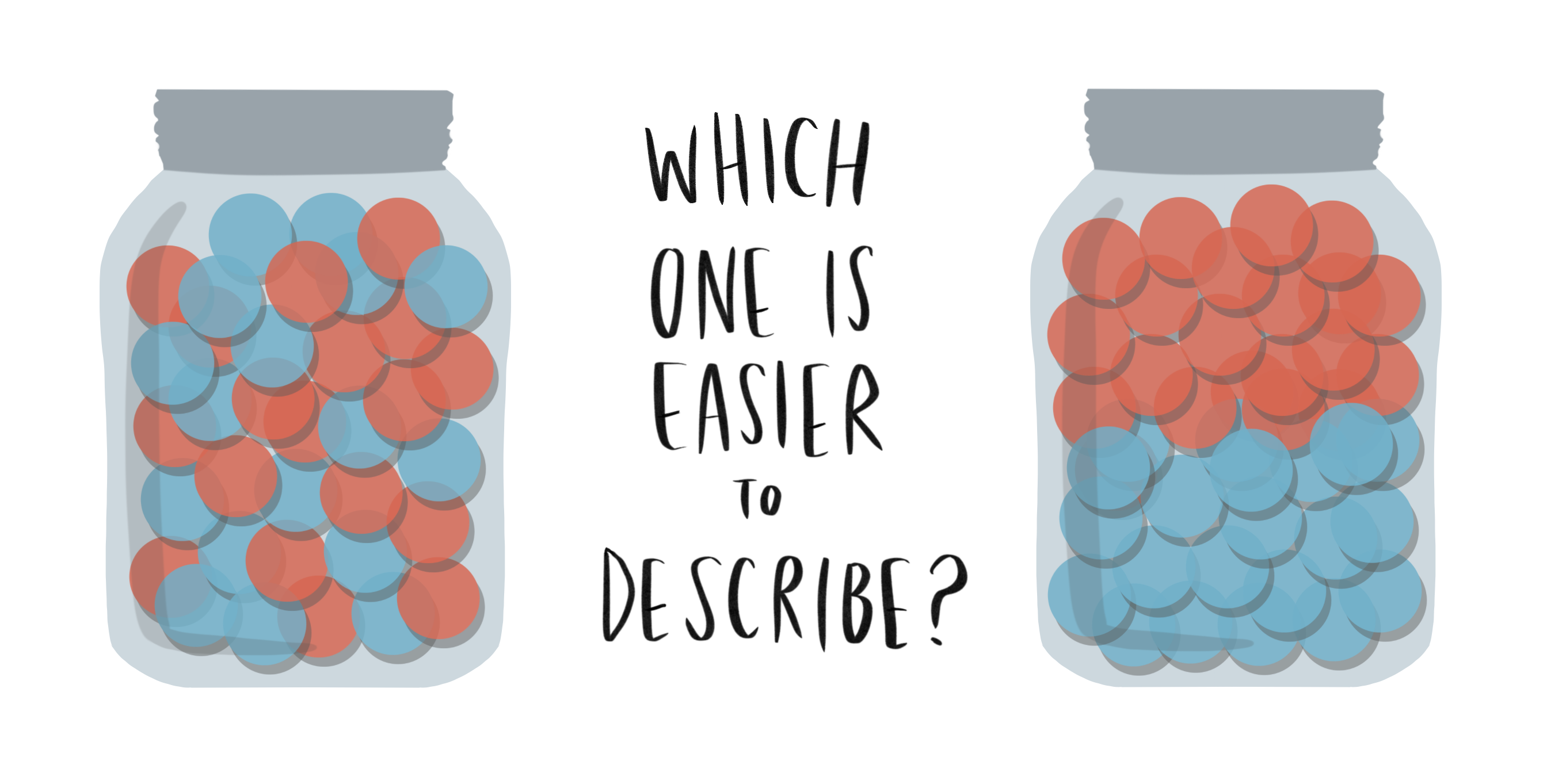}
\end{center}

In order to make his notion of entropy precise, Shannon grounded his theory on the idea that information is conveyed by binary bits. For those interested in the precise mathematical definition in the appendix to this chapter, this is where all the $\log_2$ symbols come from. For our applications, it will be enough to think of data being conveyed by a series of two-way distinctions: are you Democrat or Republican, Hispanic or non-Hispanic? For facts about people which have more than one value (like their district), we should think of encoding that data into a binary string of $0$s and $1$s that a computer might use, using the most efficient possible such encoding.

We are interested in the relationship \emph{between} datasets, not just datasets themselves. One way to quantify the relationship between two types of data associated to people is to ask how much information one tells you about the other. Returning to our town hall example, consider a situation where the room is still a fifty-fifty split, but you know that Democrats tend to sit on the left side of the hall while Republicans tend to sit on the right. You can save a lot of effort by asking all the Republicans on the left to raise their hands and asking the Democrats on the right to raise theirs, while still correctly determining the party of every participant. In this case, \emph{where} an individual was told you a lot about their party affiliation, so less effort was required to transmit the information once we knew where everyone was sitting. 

The amount of information one piece of data gives you about another can be measured using the information theory concept of \emph{conditional entropy}, denoted
$$
Ent ({X}|{Y}),
$$
which is the answer to the question, ``If I know $Y$, how much additional information, on average, do I need to also know $X$?'' 
In our applications, conditional entropy will always be calculated between two partitions of a population (e.g. into racial categories or into districts) -- in this case the data $X$ and $Y$ are just which piece of each partition an individual belongs to (e.g. their race category or their district). In this chapter, we present three examples of how conditional entropy can be applied to topics in redistricting: to measure how segregated a population is, how much counties are split up by a districting plan, or how similar two districting plans are to each other. Python code relating to this chapter is available at \url{https://github.com/political-geometry/chapter-9-entropy}.

\section{Application: Measuring Segregation}
A key issue in redistricting is the question of minority representation. Segregation plays a role in minority representation in a number of ways. For example, cases involving the Voting Rights Act often hinge on showing that a certain population group is concentrated enough to secure representation if districts were drawn in a certain way (this is the first criterion in the so-called ``Gingles test'' established by the courts in \emph{Thornburg v.~Gingles}). 

We can use conditional entropy to quantify geographical segregation by any demographic quality. In this section, we will focus on racial segregation. The segregation score we will end up with was introduced by Theil and Finizza in 1971 to study racial integration in school districts \cite{TheilFinizza} \footnote{In fact, as a quirk of the way we derive the score, the score we come up with is technically equal to one minus Theil and Finizza's score.}. Let $T$ be a partition of a geographic area into smaller units, and let $R$ be the partition of voters by race. The quantity $\re{R}{T}$ will tell us to what extent knowing where a voter resides predicts their race. However, for this measure to be comparable across different regions, we must account for the relative size of minority populations -- if the overall population is largely homogenous, it's easy to guess a voter's race even if there is significant geographic segregation. 

\begin{figure}[!]
\centering
\begin{tikzpicture}
\foreach \x in {2.75,3.75}
{
\draw[fill=red!20!white] (0.03, \x +.03) rectangle +(.94,.94); 
\draw[fill=red!20!white] (1.03, \x +.03) rectangle +(.94,.94); 
\draw[fill=blue!20!white] (2.03, \x +.03) rectangle +(.94,.94); 
\draw[fill=blue!20!white] (3.03, \x +.03) rectangle +(.94,.94); 
}

\foreach \x in {2.75,3.75}
{
\draw[fill=red!20!white] (5.03, \x +.03) rectangle +(.94,.94); 
\draw[fill=yellow!20!white] (6.03, \x +.03) rectangle +(.94,.94); 
\draw[fill=green!20!white] (7.03, \x +.03) rectangle +(.94,.94); 
\draw[fill=blue!20!white] (8.03, \x +.03) rectangle +(.94,.94); 
}

\draw[fill=red!20!white] (2.53, .03) rectangle +(.94,.94); 
\draw[fill=yellow!20!white] (3.53, .03) rectangle +(.94,.94); 
\draw[fill=green!20!white] (4.53, .03) rectangle +(.94,.94); 
\draw[fill=blue!20!white] (5.53, .03) rectangle +(.94,.94); 
\draw[fill=gray!20!white] (2.53, 1.03) rectangle +(.94,.94); 
\draw[fill=violet!20!white] (3.53, 1.03) rectangle +(.94,.94); 
\draw[fill=black!20!white] (4.53, 1.03) rectangle +(.94,.94); 
\draw[fill=cyan!20!white] (5.53, 1.03) rectangle +(.94,.94);

\end{tikzpicture}
\caption{Partitions with absolute entropy 1, 2, and 3, respectively.}
\label{absolute}
\end{figure}
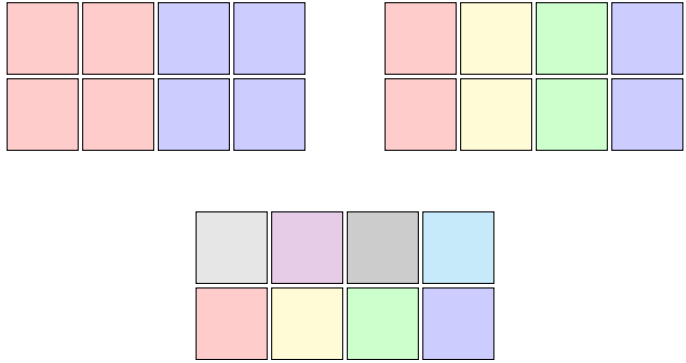

We therefore define the {\em absolute entropy} Ent${(X)}$ of a partition $X$ to be the average amount of information we need to know $X$, measured in bits. For example, if $X$ is a partition with $2^k$ parts of equal size, then Ent${(X)} = k$, as shown in Figure \ref{absolute} (this is a consequence of measuring information in binary bits, as discussed in the introduction). Now we can define an overall measure of segregation: 
$$Seg({R},{T}) = \frac{Ent({R}|{T})}{Ent{(R})}.$$

Note that Ent$({R}|{T})$ ranges from $0$ to Ent$({R})$, which explains our choice of this normalization factor. (The real-world interpretation of this upper bound is that knowing a voter's location cannot make it any harder to guess their race.) Therefore, Seg$({R},{T})$ must always be between 0 and 1. If Seg$({R},{T}) = 0$, then Ent${(R}|{T)} = 0$, which means that knowing a voter's location is completely sufficient to determine their race -- in other words, complete racial segregation. If Seg$({R},{T}) = 1$, then Ent${(R}|{T)}$ = Ent${(R)}$, so the knowledge of a voter's residence has no effect on knowledge of their race. In this case, every geographic unit has the same proportion of voters of each race, and there is no geographical segregation. 

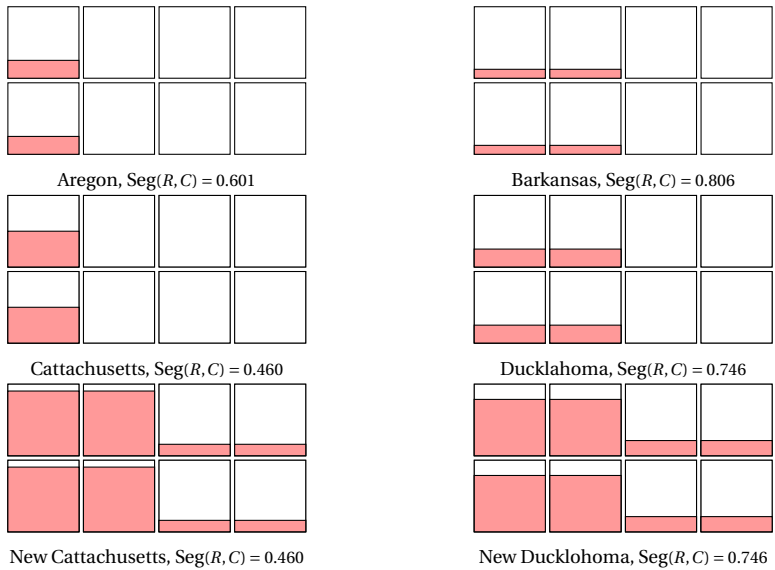
\begin{figure}[ht]
\centering
\begin{subfigure}{0.48 \textwidth}
\centering
\begin{tikzpicture}
\foreach \x in {0,1}
{
\draw[fill=white] (0.03, \x +.03) rectangle +(.94,.94); 
\draw[fill=red!40!white] (0.03, \x +.03) rectangle +(.94,0.235); 
\draw[fill=white] (1.03, \x +.03) rectangle +(.94,.94); 
\draw[fill=white] (2.03, \x +.03) rectangle +(.94,.94); 
\draw[fill=white] (3.03, \x +.03) rectangle +(.94,.94); 
}
\end{tikzpicture}
\caption*{Aregon, Seg$({R},{C})= 0.601$}
\end{subfigure}
\begin{subfigure}{0.48 \textwidth}
\centering
\begin{tikzpicture}
\foreach \x in {0,1}
{
\draw[fill=white] (5.03, \x +.03) rectangle +(.94,.94); 
\draw[fill=red!40!white] (5.03, \x +.03) rectangle +(.94,.1175); 
\draw[fill=white] (6.03, \x +.03) rectangle +(.94,.94); 
\draw[fill=red!40!white] (6.03, \x +.03) rectangle +(.94,.1175); 
\draw[fill=white] (7.03, \x +.03) rectangle +(.94,.94); 
\draw[fill=white] (8.03, \x +.03) rectangle +(.94,.94); 
}
\end{tikzpicture}
\caption*{Barkansas, Seg$({R},{C})= 0.806$}
\end{subfigure}

\begin{subfigure}{0.48 \textwidth}
\centering
\begin{tikzpicture}
\foreach \x in {0,1}
\foreach \x in {0,1}
{
\draw[fill=white] (5.03, \x +.03) rectangle +(.94,.94); 
\draw[fill=red!40!white] (5.03, \x +.03) rectangle +(.94,.47); 
\draw[fill=white] (6.03, \x +.03) rectangle +(.94,.94); 
\draw[fill=white] (7.03, \x +.03) rectangle +(.94,.94); 
\draw[fill=white] (8.03, \x +.03) rectangle +(.94,.94); 

}
\end{tikzpicture}
\caption*{Cattachusetts, Seg$({R},{C}) = 0.460$}
\end{subfigure}
\begin{subfigure}{0.48 \textwidth}
\centering
\begin{tikzpicture}
\foreach \x in {0,1}
\foreach \x in {0,1}
{
\draw[fill=white] (5.03, \x +.03) rectangle +(.94,.94); 
\draw[fill=red!40!white] (5.03, \x +.03) rectangle +(.94,.235); 
\draw[fill=white] (6.03, \x +.03) rectangle +(.94,.94); 
\draw[fill=red!40!white] (6.03, \x +.03) rectangle +(.94,.235); 
\draw[fill=white] (7.03, \x +.03) rectangle +(.94,.94); 
\draw[fill=white] (8.03, \x +.03) rectangle +(.94,.94); 

}
\end{tikzpicture}
\caption*{Ducklahoma, Seg$({R},{C})= 0.746$}
\end{subfigure}

\begin{subfigure}{0.48 \textwidth}
\centering
\begin{tikzpicture}
\foreach \x in {0,1}
\foreach \x in {0,1}
{
\draw[fill=white] (5.03, \x +.03) rectangle +(.94,.94); 
\draw[fill=red!40!white] (5.03, \x +.03) rectangle +(.94,.85); 
\draw[fill=white] (6.03, \x +.03) rectangle +(.94,.94); 
\draw[fill=red!40!white] (6.03, \x +.03) rectangle +(.94,.85); 
\draw[fill=white] (7.03, \x +.03) rectangle +(.94,.94); 
\draw[fill=red!40!white] (7.03, \x +.03) rectangle +(.94,.15); 
\draw[fill=white] (8.03, \x +.03) rectangle +(.94,.94); 
\draw[fill=red!40!white] (8.03, \x +.03) rectangle +(.94,.15); 

}
\end{tikzpicture}
\caption*{New Cattachusetts, Seg$({R},{C})= 0.460$}
\end{subfigure}
\begin{subfigure}{0.48 \textwidth}
\centering
\begin{tikzpicture}
\foreach \x in {0,1}
\foreach \x in {0,1}
{
\draw[fill=white] (5.03, \x +.03) rectangle +(.94,.94); 
\draw[fill=red!40!white] (5.03, \x +.03) rectangle +(.94,.74); 
\draw[fill=white] (6.03, \x +.03) rectangle +(.94,.94); 
\draw[fill=red!40!white] (6.03, \x +.03) rectangle +(.94,.74); 
\draw[fill=white] (7.03, \x +.03) rectangle +(.94,.94); 
\draw[fill=red!40!white] (7.03, \x +.03) rectangle +(.94,.199); 
\draw[fill=white] (8.03, \x +.03) rectangle +(.94,.94); 
\draw[fill=red!40!white] (8.03, \x +.03) rectangle +(.94,.199); 

}
\end{tikzpicture}
\caption*{New Ducklohoma, Seg$({R},{C})= 0.746$}
\end{subfigure}

\caption{Four imaginary states and their segregation scores}
\label{seg}
\end{figure}

To try and get an idea of how our score behaves, we consider four fictional states -- Aregon, Barkansas, Cattachusetts and Ducklahoma, shown in Figure \ref{seg}. Aregon and Barkansas each have a statewide minority population share of 1/16th, while Cattachusetts and Ducklahoma each have a statewide minority share of 1/8th. Aregon is visibly more segregated than both Barkansas and Ducklahoma, and the scores indicate this. On the other hand, Aregon could be considered less segregated than Cattachusetts because despite an increase in minority population, the region to which the minority population is confined has not increased; the scores agree with this intuition. We also want to investigate how two states with the same segregation score but with vastly different statewide minority populations can look. We thus construct two more states, New Cattachusetts and New Ducklahoma, each designed to have the same score as their predecessors but to have equal pink and white populations. While the difference in segregation between Cattachusetts and Ducklahoma is obvious to the eye, the difference between New Cattachusetts and New Ducklahoma is more subtle despite having the same score difference. This serves as a cautionary tale about interpreting the absolute difference between segregation scores, which will be important to keep in mind when we move to real-world data. 

\begin{figure}[!ht]
\centering
\begin{subfigure}{0.32\textwidth}
\centering
\includegraphics[width=\textwidth]{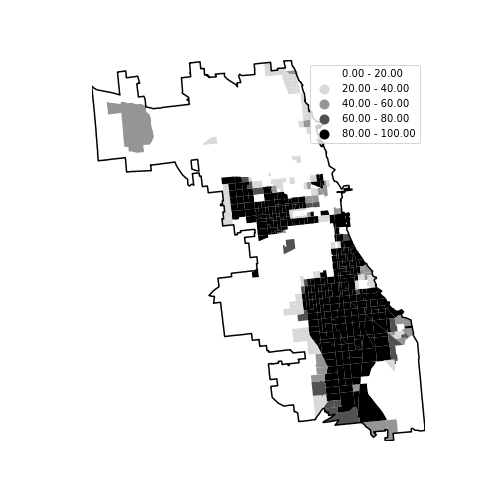}
\caption*{1990, Black}
\end{subfigure}
\begin{subfigure}{0.32\textwidth}
\centering
\includegraphics[width=\textwidth]{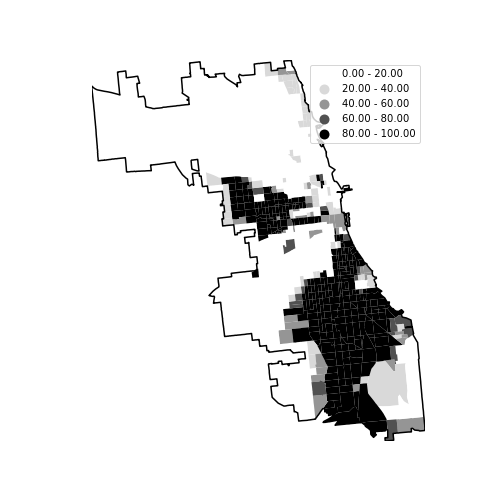}
\caption*{2000, Black}
\end{subfigure}
\begin{subfigure}{0.32\textwidth}
\centering
\includegraphics[width=\textwidth]{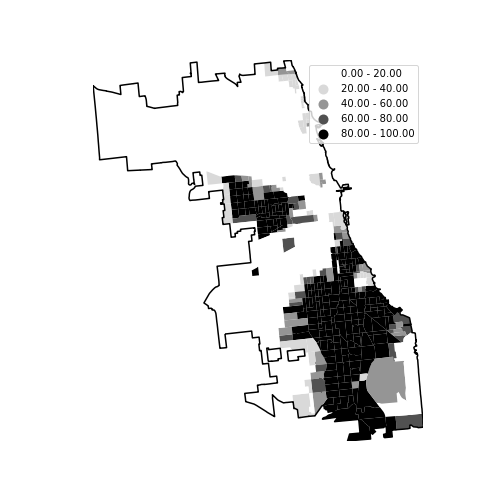}
\caption*{2010, Black}
\end{subfigure}

\begin{subfigure}{0.32\textwidth}
\centering
\includegraphics[width=\textwidth]{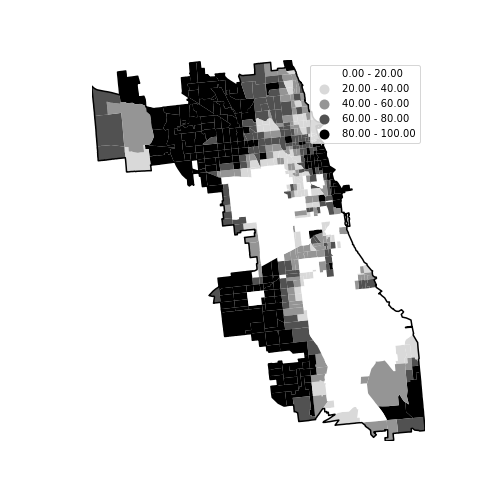}
\caption*{1990, White}
\end{subfigure}
\begin{subfigure}{0.32\textwidth}
\centering
\includegraphics[width=\textwidth]{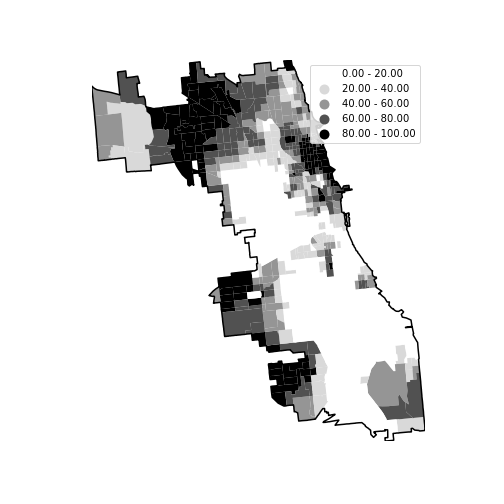}
\caption*{2000, White}
\end{subfigure}
\begin{subfigure}{0.32\textwidth}
\centering
\includegraphics[width=\textwidth]{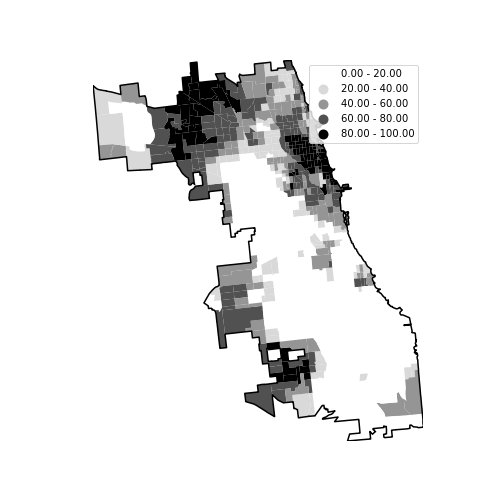}
\caption*{2010, White}
\end{subfigure}

\begin{subfigure}{0.32\textwidth}
\centering
\includegraphics[width=\textwidth]{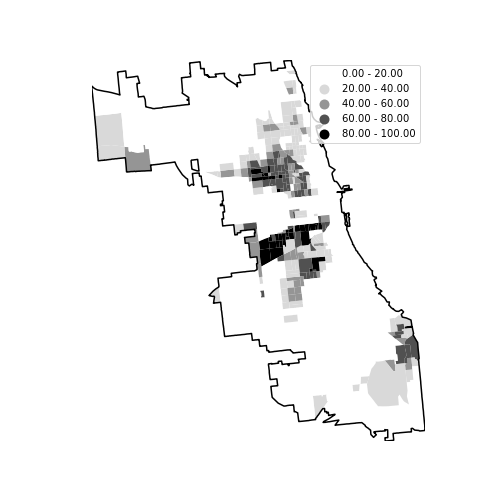}
\caption*{1990, Hispanic}
\end{subfigure}
\begin{subfigure}{0.32\textwidth}
\centering
\includegraphics[width=\textwidth]{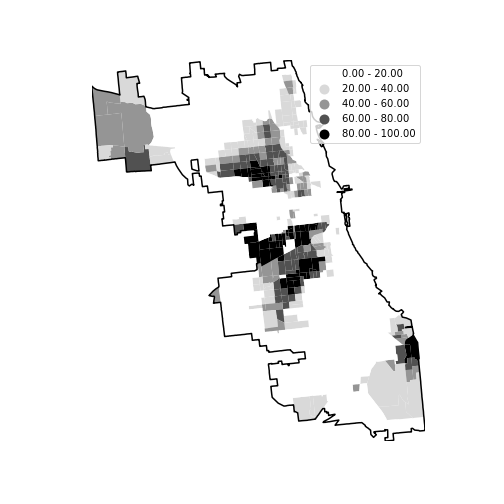}
\caption*{2000, Hispanic}
\end{subfigure}
\begin{subfigure}{0.32\textwidth}
\centering
\includegraphics[width=\textwidth]{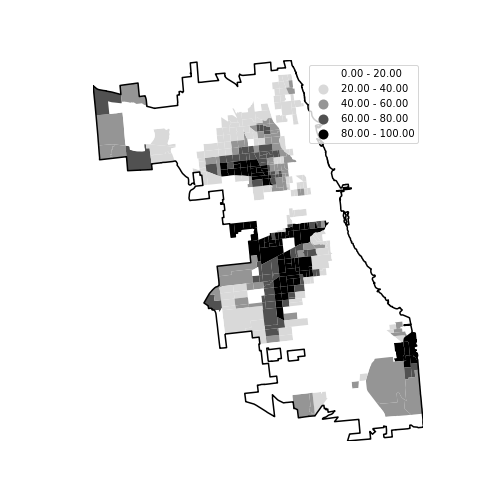}
\caption*{2010, Hispanic}
\end{subfigure}
\caption{Race category choropleths for Chicago 1990--2010}
\label{fig:choro}
\end{figure}

To illustrate this measure on a real-world example, we calculate racial segregation scores in the city of Chicago, using Census demographic data from 1990, 2000, and 2010 \footnote{In this survey, Hispanic is considered an ethnicity and can be selected independently of race variables like Black and White. For our calculations, we therefore categorize as Hispanic all persons who selected a Hispanic ethnicity, and denote as Black and White only people who selected non-Hispanic ethnicities and a single race category. Also, since some census tracts may cross the city boundary, we take as our domain of study all tracts which intersect the city.} For each category, we report the segregation Seg$({R},{T})$ where $T$ is the partition of the residents of Chicago into census tracts and $R$ is the bipartition given by whether or not residents are in the specified category. Table \ref{table:chitown} gives the results. The maps in Figure \ref{fig:choro} show the racial demographics of the Chicago tracts for comparison (such maps are called choropleths).

\begin{table}
\centering
\begin{tabular}{| l | c | c | c |}
\hline
Year & Black & White & Hispanic \\ \hline
1990 & 0.262 & 0.514 & 0.596 \\
2000 & 0.315 & 0.589 & 0.591 \\
2010 & 0.377 & 0.625 & 0.610  \\
\hline
\end{tabular}
\caption{Segregation score by race category and year for Chicago}
\label{table:chitown}
\end{table}

\begin{table}
\centering
\begin{tabular}{| l |c|c|c|}
\hline
Year & Black & White & Hispanic \\ \hline
1990 & 35.7\% & 41.7\% & 18.9\% \\
2000 & 34.0\% & 34.3\% & 25.5\% \\
2010 & 30.2\% & 34.1\% & 28.7\%  \\
\hline
\end{tabular}
\caption{City-wide racial demographics of Chicago census tracts}
\label{citydemos}
\end{table}

Recall that a higher score indicates less segregation. For all the years considered, Black/non-Black segregation was higher than Hispanic/non-Hispanic -- in information theory terms, knowing where someone resides improves one's chances of guessing whether they are Black or not more than it improves one's chances of guessing whether they are Hispanic or not. Table \ref{table:chitown} seems to indicate that the Black/non-Black and White/non-White segregation has decreased over time, while the segregation of the Hispanic population has remained fairly steady, decreasing only slightly from 1990 to 2010. The decrease in White/non-White segregation is observable from the choropleths in Figure \ref{fig:choro}, while the decrease in Black/non-Black segregation is less apparent. The number of census tracts with less than 1\% Black population went from 252 in 1990, to 165 in 2000 and to 91 in 2010 (each out of about one thousand total census tracts), which gives some evidence for decreased Black/non-Black segregation. The difficulty of detecting this on the choropleths is a good motivation for a precise score such as Seg$({R},{T})$ to measure segregation. As for the Hispanic population, both the choropleths and entropy scores indicate that despite a significant increase in population overall (from 19\% in 1990 to 29\% in 2010), the Hispanic/non-Hispanic segregation has persisted to a large degree.

An important feature of the entropy segregation score is that it is highly dependent on the geographic units chosen -- in this case, census tracts. Indeed, for the finest units imaginable -- individual persons -- knowing which geographic unit someone belongs to completely determines that person's race. At the coarsest scale, knowing that someone is in Chicago does not give you any information about their race beyond that given by the city-wide demographics, i.e., the data captured by Ent$({R})$. Using this score thus requires a careful choice of particular \emph{scale} at which one is trying to detect segregation. This is a downside if one is interested in a scale-free measure, but it can also be an advantage. For example, in congressional redistricting one may be more concerned about a coarser scale of segregation than when dealing with public policy at the level of a single city.

Measures of segregation have an extensive history, which we will not be able to cover in this section -- see for example \cite{MasseyDenton} for a classic survey of some 20 scores, including the entropy score studied in this section. Some of these scores, such as the Dissimilarity score, the Frey index and the segregation Gini index, are calculated using a specific choice of geographic unit, but beyond that do not take geography (e.g. adjacency of units) into account, just like the entropy score in this section. The most widely used measure which does take geographic adjacency into account is Moran's I introduced in 1950 by P. A. P. Moran. However, Moran's I still depends heavily on the geographic units chosen. For a modern method based on network science which is less sensitive to choice of geographic unit, see \cite{capy}.

\section{Application: Splitting Counties}

Many state constitutions prohibit voting districts from dividing up counties. In cases when counties are too populous to completely avoid splitting, there is often a legal requirement to minimize the amount of division. Sometimes this is measured by the number of counties split, but in other cases there is no specified metric. 

Conditional entropy gives us a precise and meaningful way to quantify the extent to which a districting plan $D$ divides up the county partition $C$. The quantity Ent$({D}|{C})$ tells us how much we can deduce about a voter's district given their county of residence. If this quantity is low, then counties are not split very much; if it is high, they are split drastically.

In this simple example, our hypothetical state consists of four counties of equal size, denoted with colors. We will calculate the conditional entropy for several different districting plans.

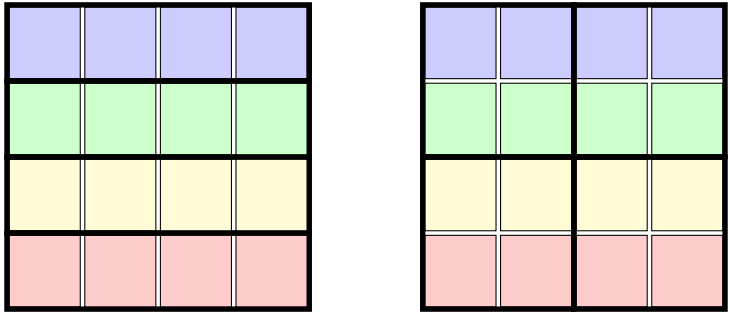
\begin{figure}
\centering
\begin{tikzpicture}
\foreach \x in {0,...,3}
{
\draw[fill=red!20!white] (\x +.03,0.03) rectangle +(.94,.94); 
\draw[fill=yellow!20!white] (\x +.03,1.03) rectangle +(.94,.94); 
\draw[fill=green!20!white] (\x +.03,2.03) rectangle +(.94,.94); 
\draw[fill=blue!20!white] (\x +.03,3.03) rectangle +(.94,.94); 
\draw[black, line width=2pt] (0, \x) rectangle +(4,1);
}

\foreach \x in {5.5,...,8.5}
{
\draw[fill=red!20!white] (\x +.03,0.03) rectangle +(.94,.94); 
\draw[fill=yellow!20!white] (\x +.03,1.03) rectangle +(.94,.94); 
\draw[fill=green!20!white] (\x +.03,2.03) rectangle +(.94,.94); 
\draw[fill=blue!20!white] (\x +.03,3.03) rectangle +(.94,.94); 
}
\draw[black, line width=2pt] (5.5, 0) rectangle +(2,2);
\draw[black, line width=2pt] (5.5, 2) rectangle +(2,2);
\draw[black, line width=2pt] (7.5, 0) rectangle +(2,2);
\draw[black, line width=2pt] (7.5, 2) rectangle +(2,2);
\end{tikzpicture}
\caption{The left-hand plan splits no counties and has a conditional entropy of 
Ent$({D}|{C})=0$. The right-hand plan bisects each county and has conditional entropy Ent$({D}|{C})=1$.}
\label{entzeroone}
\end{figure}

Consider one extreme case, shown on the left in Figure \ref{entzeroone}, where no counties are split. In this case, knowing a voter's county fully determines their voting district, and therefore no additional information is needed. In this case, Ent$({D}|{C}) = 0$. 

Suppose, however, that our plan is not able to do so well. Take the case where each county is divided evenly into two districts, as in the right side of Figure \ref{entzeroone}. Then, if we know which county a voter resides in, we can narrow down their district into two options. We would need exactly one more binary bit of information (i.e. one more two-way distinction) to determine their district. In this case, Ent$({D}|{C})= 1$.

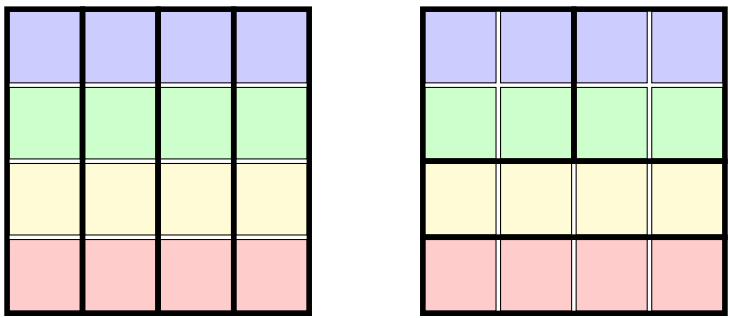
\begin{figure}
\centering
\begin{tikzpicture}
\foreach \x in {0,...,3}
{
\draw[fill=red!20!white] (\x +.03,0.03) rectangle +(.94,.94); 
\draw[fill=yellow!20!white] (\x +.03,1.03) rectangle +(.94,.94); 
\draw[fill=green!20!white] (\x +.03,2.03) rectangle +(.94,.94); 
\draw[fill=blue!20!white] (\x +.03,3.03) rectangle +(.94,.94); 
\draw[black, line width=2pt] (\x, 0) rectangle +(1,4);
}

\foreach \x in {5.5,...,8.5}
{
\draw[fill=red!20!white] (\x +.03,0.03) rectangle +(.94,.94); 
\draw[fill=yellow!20!white] (\x +.03,1.03) rectangle +(.94,.94); 
\draw[fill=green!20!white] (\x +.03,2.03) rectangle +(.94,.94); 
\draw[fill=blue!20!white] (\x +.03,3.03) rectangle +(.94,.94); 
}
\draw[black, line width=2pt] (5.5, 0) rectangle +(4,1);
\draw[black, line width=2pt] (5.5, 2) rectangle +(2,2);
\draw[black, line width=2pt] (5.5, 1) rectangle +(4,1);
\draw[black, line width=2pt] (7.5, 2) rectangle +(2,2);
\end{tikzpicture}
\caption{For a districting plan that cuts each county into four equal pieces, we get  Ent$({D}|{C})=2$. The plan on the right is mixed, but averages out to a conditional entropy of Ent$({D}|{C})=0.5$.}
\label{enttwo}
\end{figure}

To take the opposite extreme, suppose all of the counties are evenly divided into four pieces by the districting plan, as shown in the left side of Figure \ref{enttwo}. In this case, knowing a voter's county tells us nothing at all about their district. We would need to be told outright which of the four districts they lived in, which requires 2 bits of information. In this case, Ent$({D}|{C})$ = 2.

Lastly, consider the plan on the right side of Figure \ref{enttwo}, which splits two counties but leaves two intact. For a voter in the blue or green county, one additional bit of information is needed to specify their district. However, for a resident of the yellow or red county, no information is needed. Thus, averaging over the entire state, we see that Ent$({D}|{C})= 0.5$.

\begin{figure}
\begin{subfigure}{0.475\textwidth}
\centering
\includegraphics[width=\textwidth]{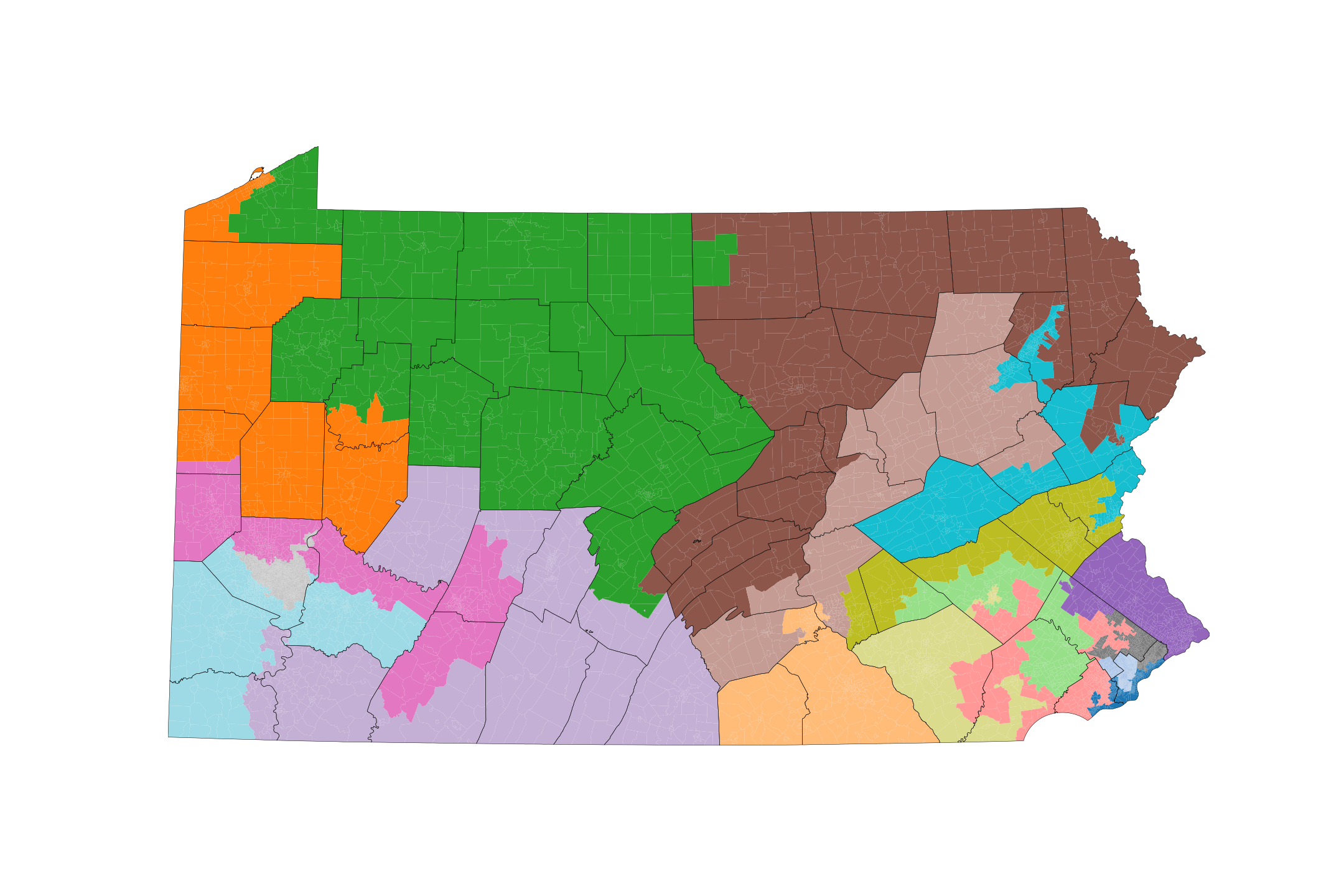}
\caption*{2011}
\end{subfigure}
\begin{subfigure}{0.475\textwidth}
\centering
\includegraphics[width=\textwidth]{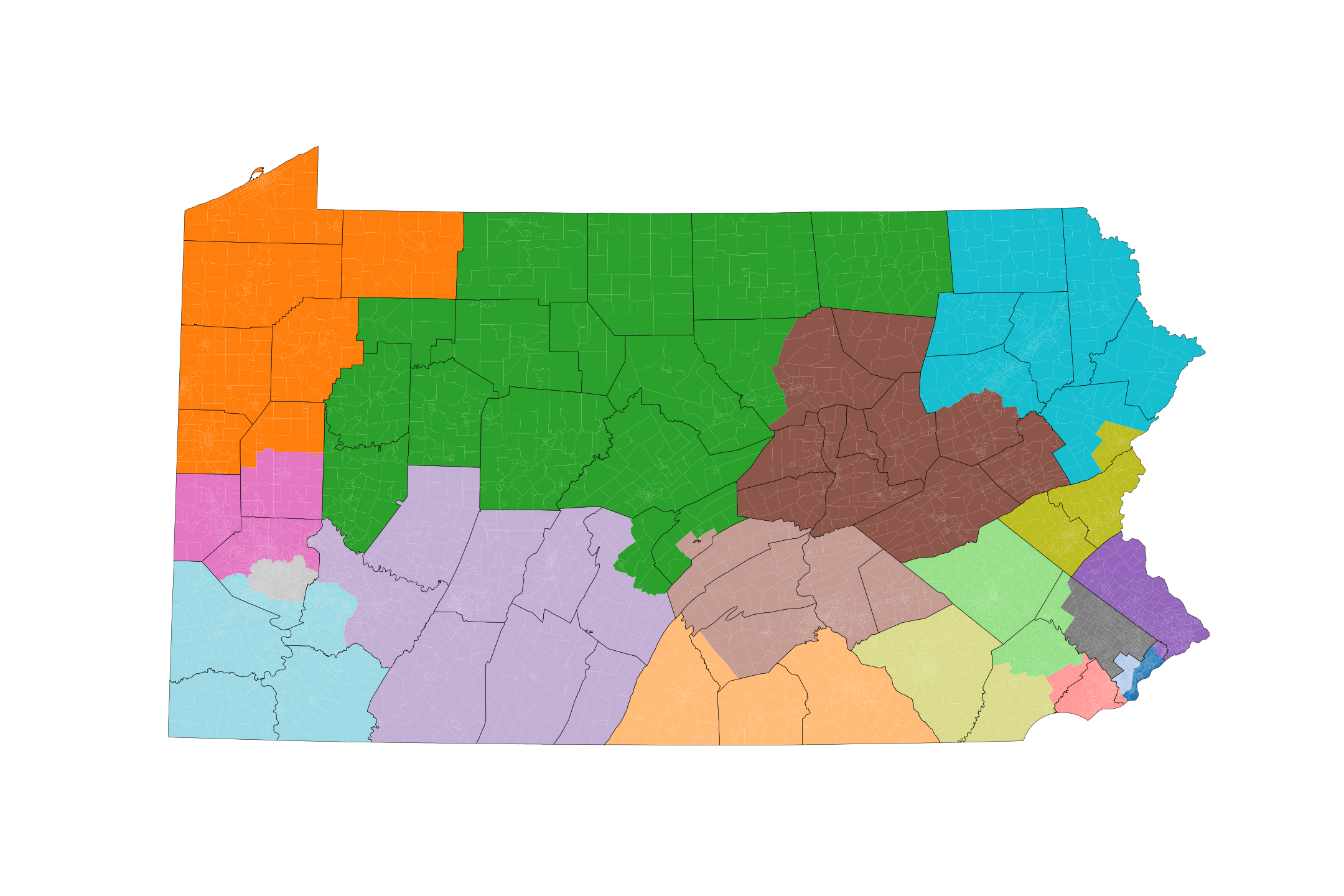}
\caption*{CPCT}
\end{subfigure}

\begin{subfigure}{0.475\textwidth}
\includegraphics[width=\textwidth]{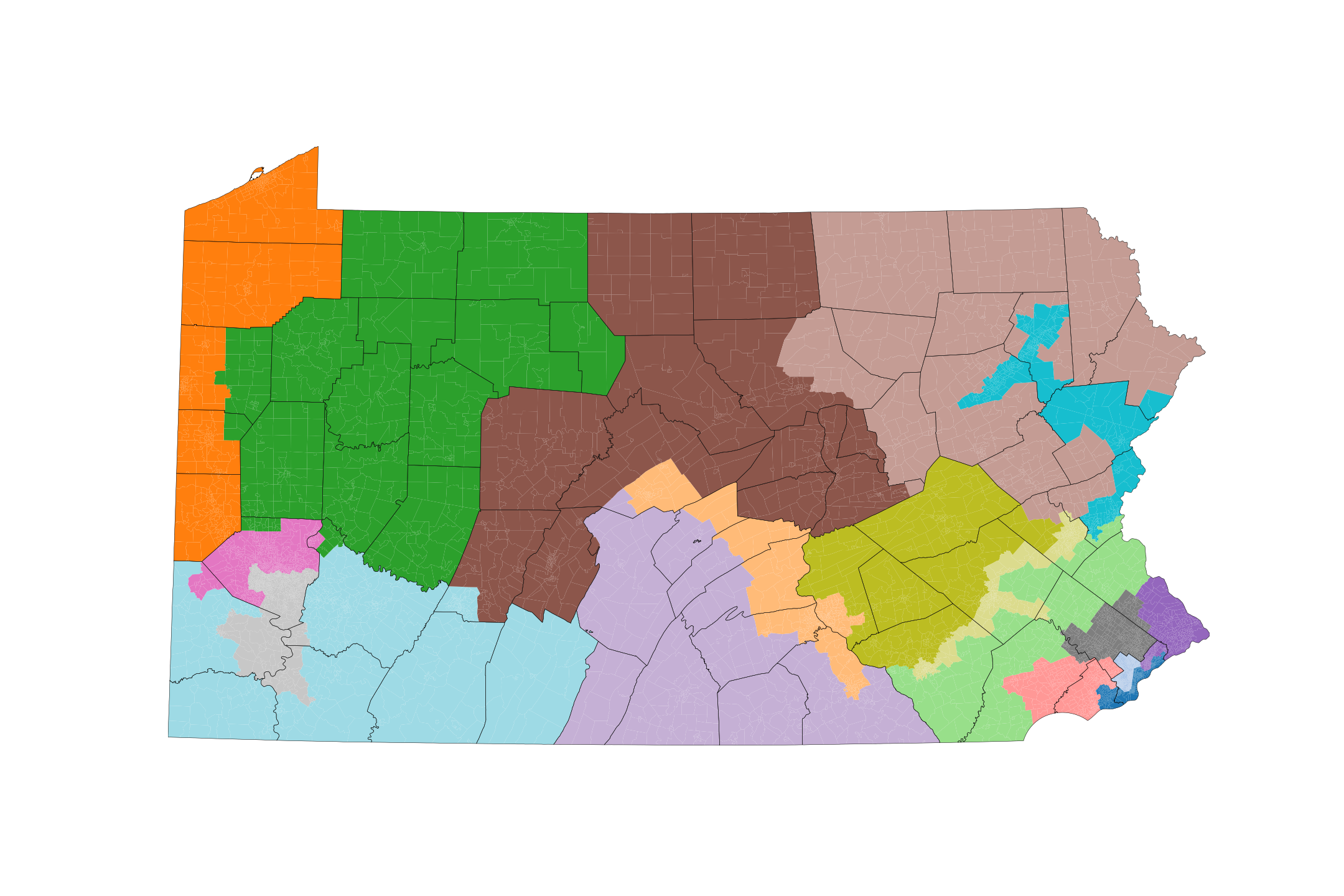}
\caption*{DEM}
\end{subfigure}
\begin{subfigure}{0.475\textwidth}
\includegraphics[width=\textwidth]{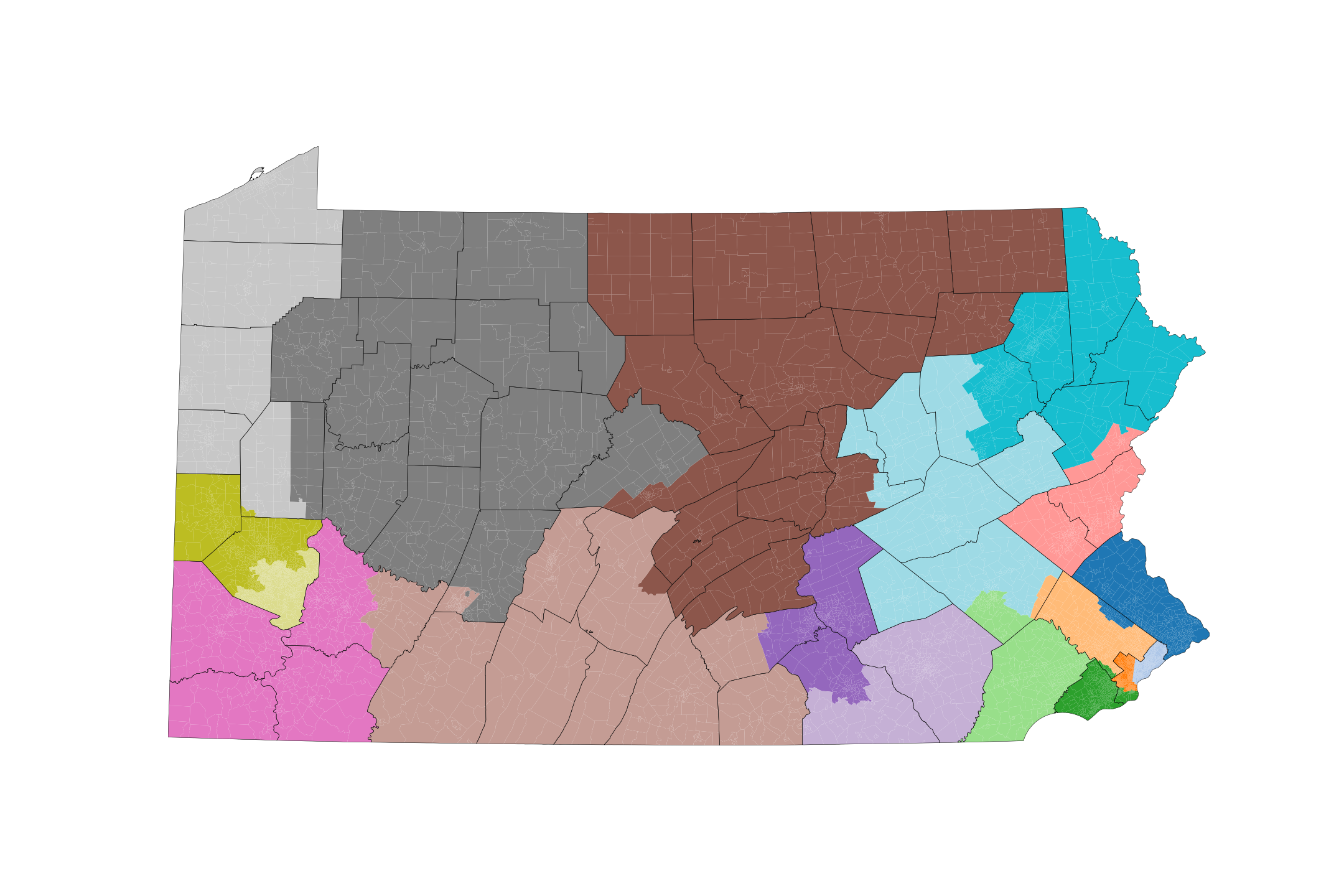}
\caption*{REMEDIAL}
\end{subfigure}

\begin{subfigure}{0.475\textwidth}
\includegraphics[width=\textwidth]{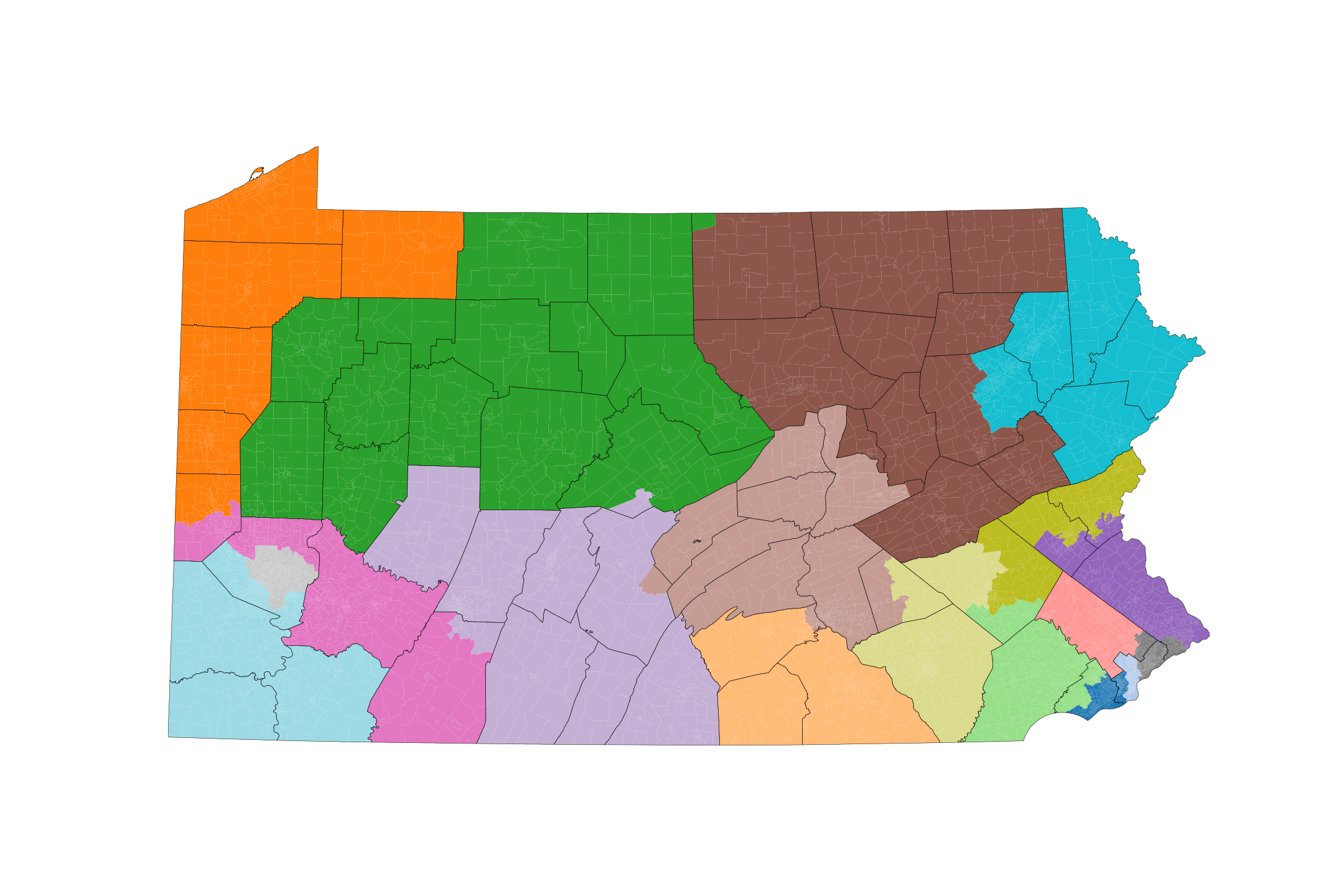}
\caption*{GOV}
\end{subfigure}
\begin{subfigure}{0.475\textwidth}
\includegraphics[width=\textwidth]{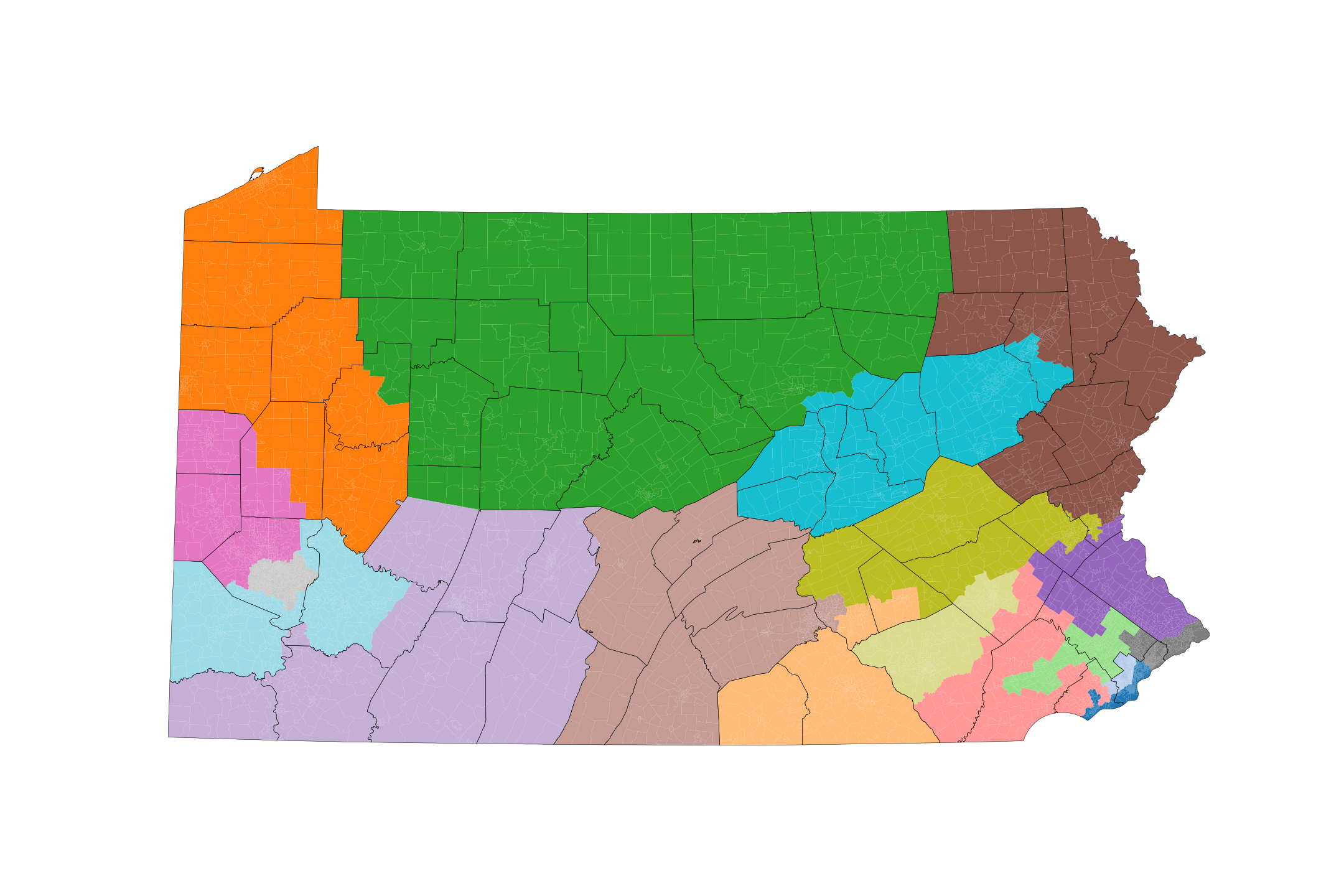}
\caption*{GOP}
\end{subfigure}

\begin{subfigure}{0.475\textwidth}
\includegraphics[width=\textwidth]{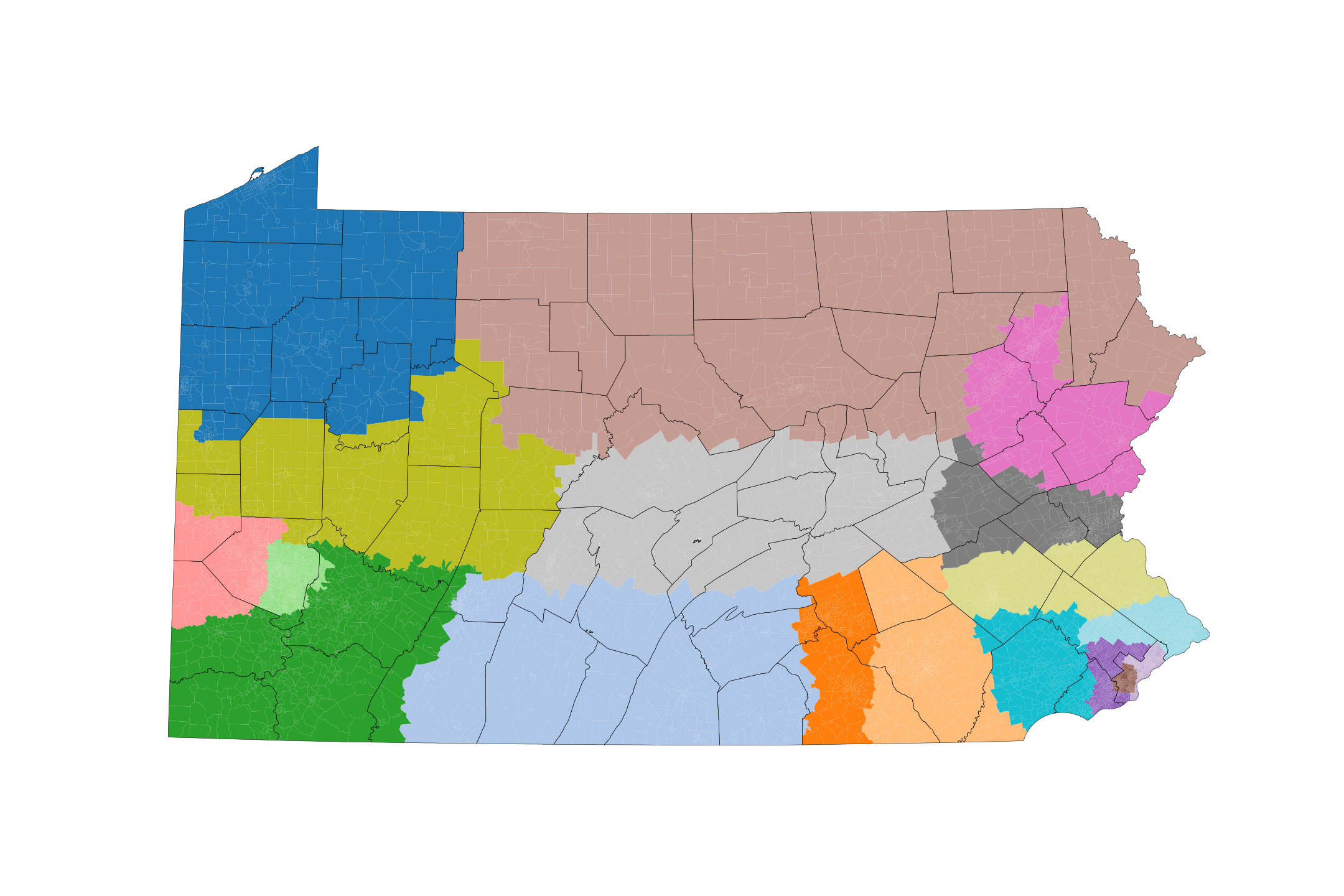}
\caption*{8th}
\end{subfigure}
\begin{subfigure}{0.475\textwidth}
\includegraphics[width=\textwidth]{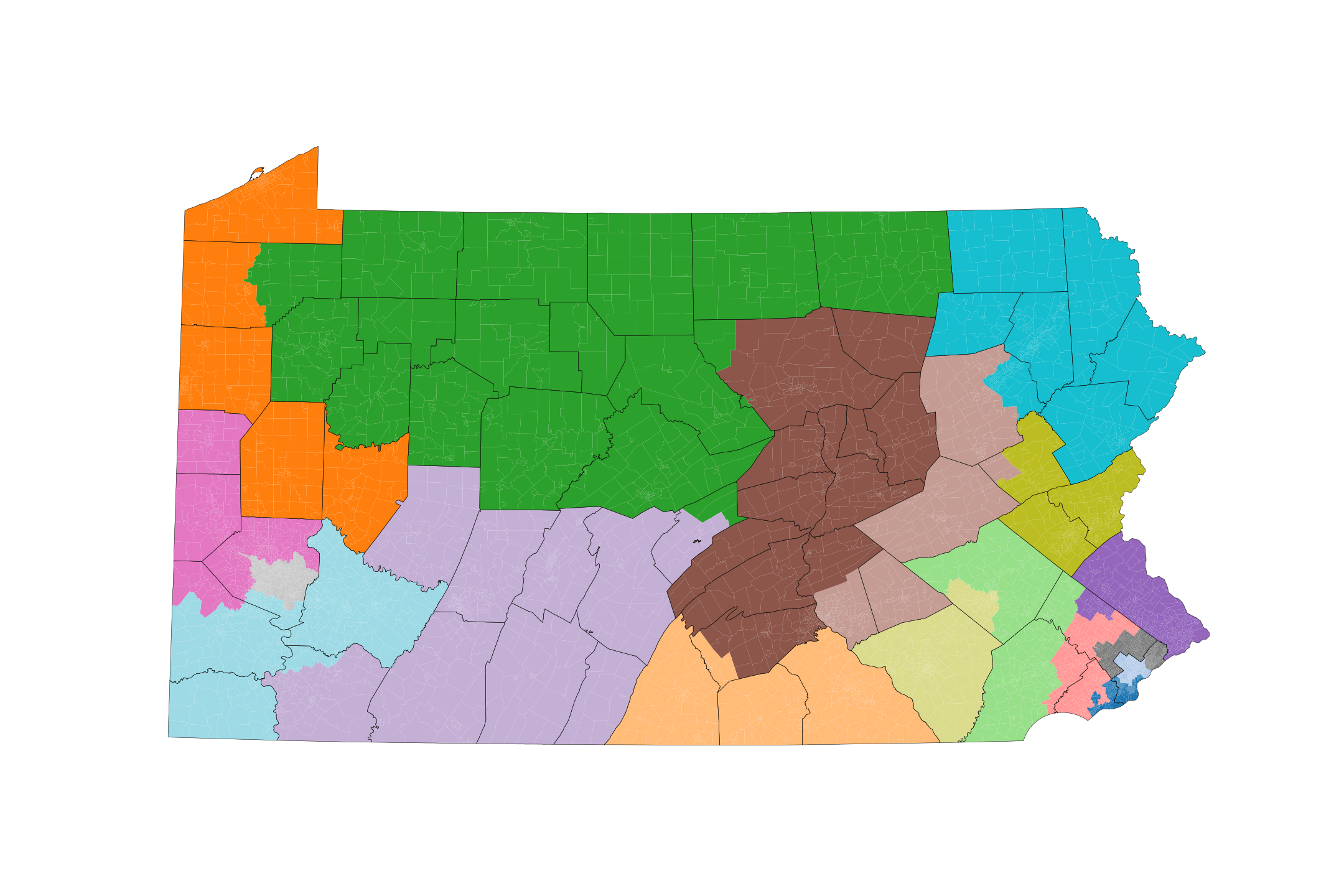}
\caption*{TS}
\end{subfigure}

\caption{Districts (shaded regions) over county boundaries (in black)}
\label{fig:PAmaps}
\end{figure}

To see how this plays out in practice, we compute Ent$({D}|{C})$ for eight different congressional districting plans for Pennsylvania drawn by various groups and individuals, as summarized below and shown in Figure \ref{fig:PAmaps}. 

\begin{itemize}
\item \emph{2011}: The congressional districting plan enacted by the Republican-controlled state legislature and signed into law by Governor Tom Corbett (R) in December 2011. The map was challenged in court and in January 2018 the Pennsylvania Supreme Court struck down the map for being an illegal partisan gerrymander, giving parties to the suit until February 15, 2018 to propose new maps to the court.
\item \emph{TS}: The plan submitted directly to Governor Tom Wolf (D) by House Speaker Mike Turzai (R) and Senate President Pro Tem Joe Scarnati (R). Governor Wolf declined to submit the map to the PA Supreme Court, saying that it was a Republican gerrymander. 
\item \emph{GOV}: The plan submitted by Governor Tom Wolf (D) to the PA Supreme Court. 
\item \emph{REM}: The remedial map drawn by Stanford Professor Nate Persily at the direction of the PA Supreme Court. This map was adopted by the court on February 19, 2018 and was still in place at the time this chapter was written. 
\item \emph{8th}: Districting plan drawn by a class of eighth-graders (who were not involved to the lawsuit around the 2011 plan).
\item Three plans from FiveThirtyEight's ``The Atlas of Redistricting'', a collection of congressional maps for every state optimizing various criteria \cite{538}.
\begin{itemize}
\item \emph{538 CPCT}: plan optimized for compactness.
\item \emph{538 DEM}: plan optimized for Democrats.
\item \emph{538 GOP}: plan optimized for Republicans.
\end{itemize} 
\end{itemize}

In addition to our entropy score, we also compute two more natural measures of county splitting for each plan:
\begin{itemize}
\item \emph{splits}, the total number of counties which intersect more than one district, and
\item \emph{pieces}, which we compute by overlaying the district boundaries and county boundaries and counting the number of connected pieces in the resulting partition of the state
\end{itemize}
Table \ref{tab:PAallscore} shows all three county splitting scores for the plans. 

While \emph{splits} cares only about the number of counties split and not how they are split (e.g. 50/50 or 99/1), entropy takes the nature of the splits into account. This can be considered an advantage or a disadvantage. On the one hand, Ent$({D}|{C})$ is a finer measure of county splitting which is grounded in a real-world question (if I know my county, how well do I know my district?). On the other hand, the entropy score barely penalizes ``nibbles'' where a district includes a tiny fraction of a county. Intuitively, this is because most of the county can still easily guess their district correctly. Whether such nibbles should be considered negligible, a significant downside or even the worst possible way to split a county depends on the perspectives of stakeholders.

In Pennsylvania, as in many states, county populations vary wildly -- from about 1.5 million in Philadelphia County to a mere 5000 or so in Cameron County. The entropy splitting score naturally up-weights more populous counties: splitting Philadelphia County in half adds 300 times more to the entropy score than splitting Cameron county.  On the other hand, \emph{splits} and \emph{pieces} treat all counties the same regardless of population. This is again a question of priorities -- but observe that Iowa's redistricting code, for example, requires that ``[w]hen there is a choice between dividing local political subdivisions, the more populous subdivisions shall be divided before the less populous'' \cite{Iowa}. 

Other ways to measure county splitting proposed in the literature include entropy-like scores such as the one proposed in footnote 6 of \cite{VApaper}, which adapts the entropy score to be more sensitive to the above-mentioned ``nibbles''. In their Markov Chain Monte Carlo sampling method, Mattingly et al. impose a penalty for counties which are split between two districts and a much larger penalty for counties which are split between three or more districts, with weight factors depending on the size of the intersections between a county and any extra districts~\cite{Mattingly}. Beyond the realm of computational methods (which are often benefited by finer, real-valued measures such as entropy), most expert reports dealing with county splitting use discrete measures such as \emph{splits}. See for example Kennedy's expert report in \emph{League of Women Voters of PA v. Pennsylvania}, which places a great deal of emphasis on the number of counties split (i.e. the \emph{splits} score) by the 2011 plan compared to its predecessors \cite{Kennedy}, or the court's emphasis on the low number of counties split in their decision to adopt the REM plan \cite{REMorder}.

\begin{table}
\centering
\begin{tabular}{|l|c|c|c|}
\hline
Plan & $\re{D}{C}$ & \emph{splits} & \emph{pieces} \\ \hline
REM & 0.474 & 14 & 85\\
538 CPCT & 0.482 & 12 & 83  \\
GOV & 0.579 & 18 & 88 \\
TS & 0.601 & 15 & 87 \\
538 GOP & 0.816 & 21 & 102 \\
8TH &  0.833 & 42 &  138\\
2011 & 0.868 & 28 & 111  \\
538 DEM & 0.920 & 31 & 117 \\
\hline
\end{tabular}
\caption{County-splitting scores for Pennsylvania plans.  \\ (\emph{splits} = \# counties split by districts, \emph{pieces} = \# connected pieces when county and district boundaries are overlaid)}
\label{tab:PAallscore}
\end{table}

\section{Application: Distance Between Plans}

In the redistricting process, it is often useful to ask the question ``how different are two districting plans?'' For example, a redistricting process may legally require one to satisfy certain constraints while also remaining as ``close'' as possible to the existing districts. Or one may want to know when a particular plan is an outlier compared with a collection of other plans, either human-drawn or generated by some algorithm.

Conditional entropy provides us with a mathematically precise answer to this question. If we have two redistricting plans $D_1$ and $D_2$, we can look at their conditional entropy $\re{D_1}{D_2}$ to see how much they coincide.\footnote{Although it is not generally true that $\re{X}{Y} = \re{Y}{X}$ (see Section \ref{formal}) it is possible to prove that $\re{D_1}{D_2} = \re{D_2}{D_1}$ so long as each plan has the same number of districts, and the districts are perfectly population-balanced. In practice we use the average of $\re{D_1}{D_2}$ and $\re{D_2}{D_1}$ when computing the distance} To get some notion of what this number means, we may look at two extreme cases. If $D_1 = D_2$, then knowing a voter's district under $D_2$ is completely sufficient to know that voter's district under $D_1$, and therefore $\re{D_1}{D_2} = 0$. On the other end, the maximum possible value of $\re{D_1}{D_2}$ is simply $\ent{D_1}$ (which itself is bounded above by $\log_2(k)$ where $k$ is the number of districts), which corresponds to the case that the two plans are as different as they could possibly be: knowing a voter's district under $D_2$ tells you nothing about their district under $D_1$, and vice versa.

\begin{figure}
\centering
\begin{tikzpicture}
\begin{scope}[node distance=1]
\node (top) {};
\coordinate[below of=top] (c);
\node[inner sep=0pt, left=of c, align=right] (MD102)
    {\includegraphics[width=0.4\textwidth]{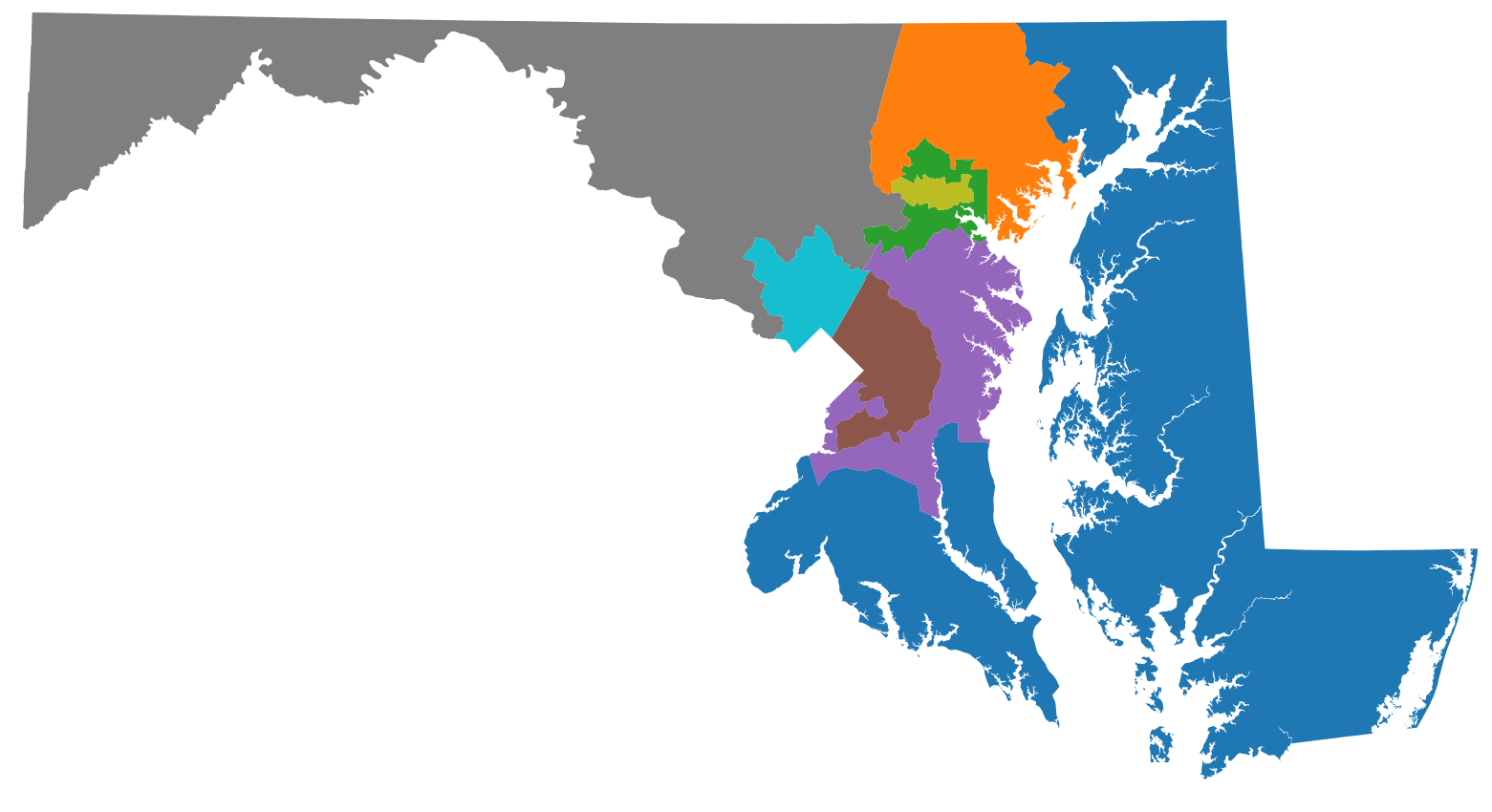}};
\node[inner sep=0pt, right=of c, align=right] (MD103)
    {\includegraphics[width=0.4\textwidth]{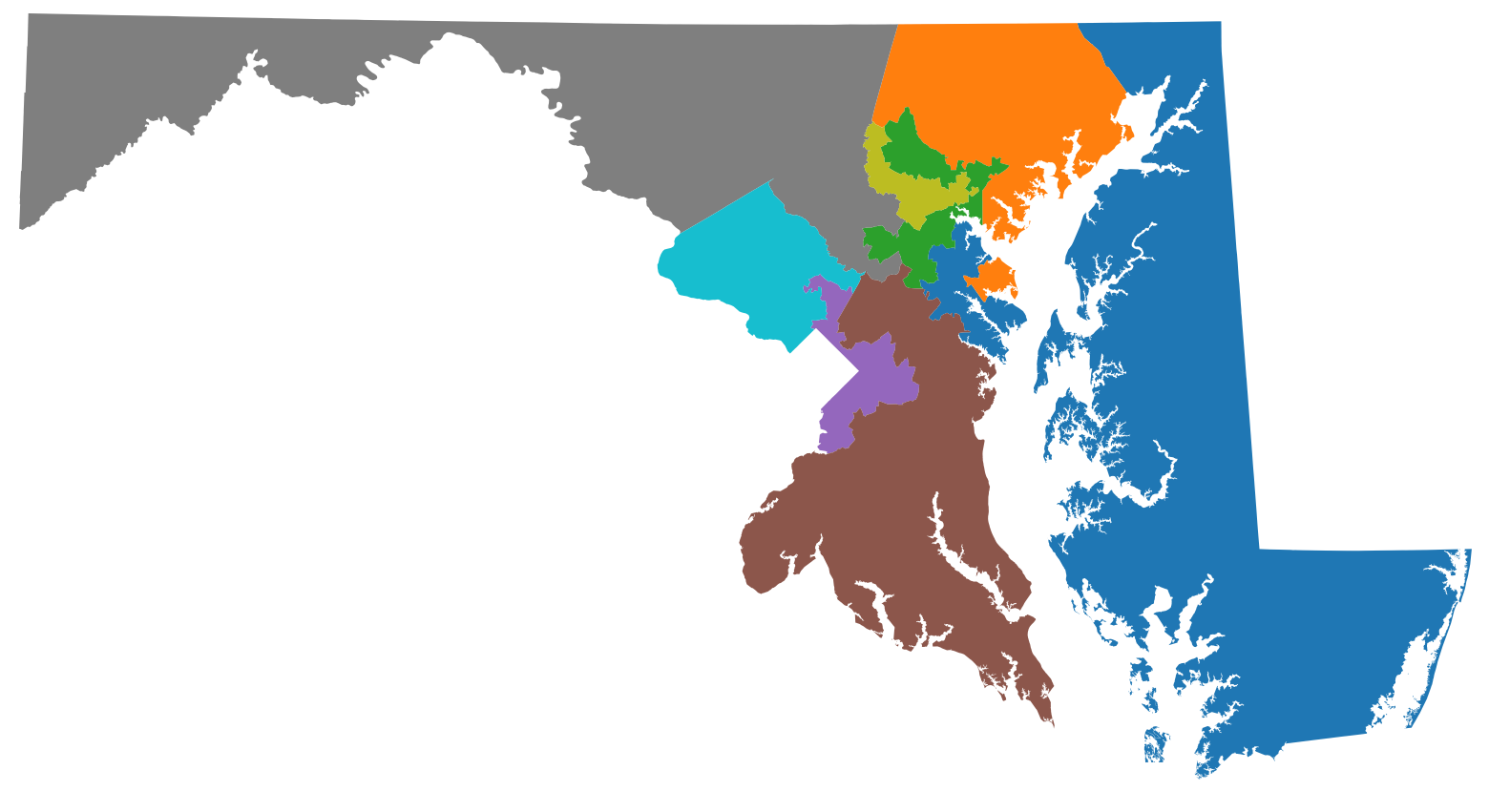}};
\node[inner sep=0pt, below=of MD102, align=right] (MD108)
    {\includegraphics[width=0.4\textwidth]{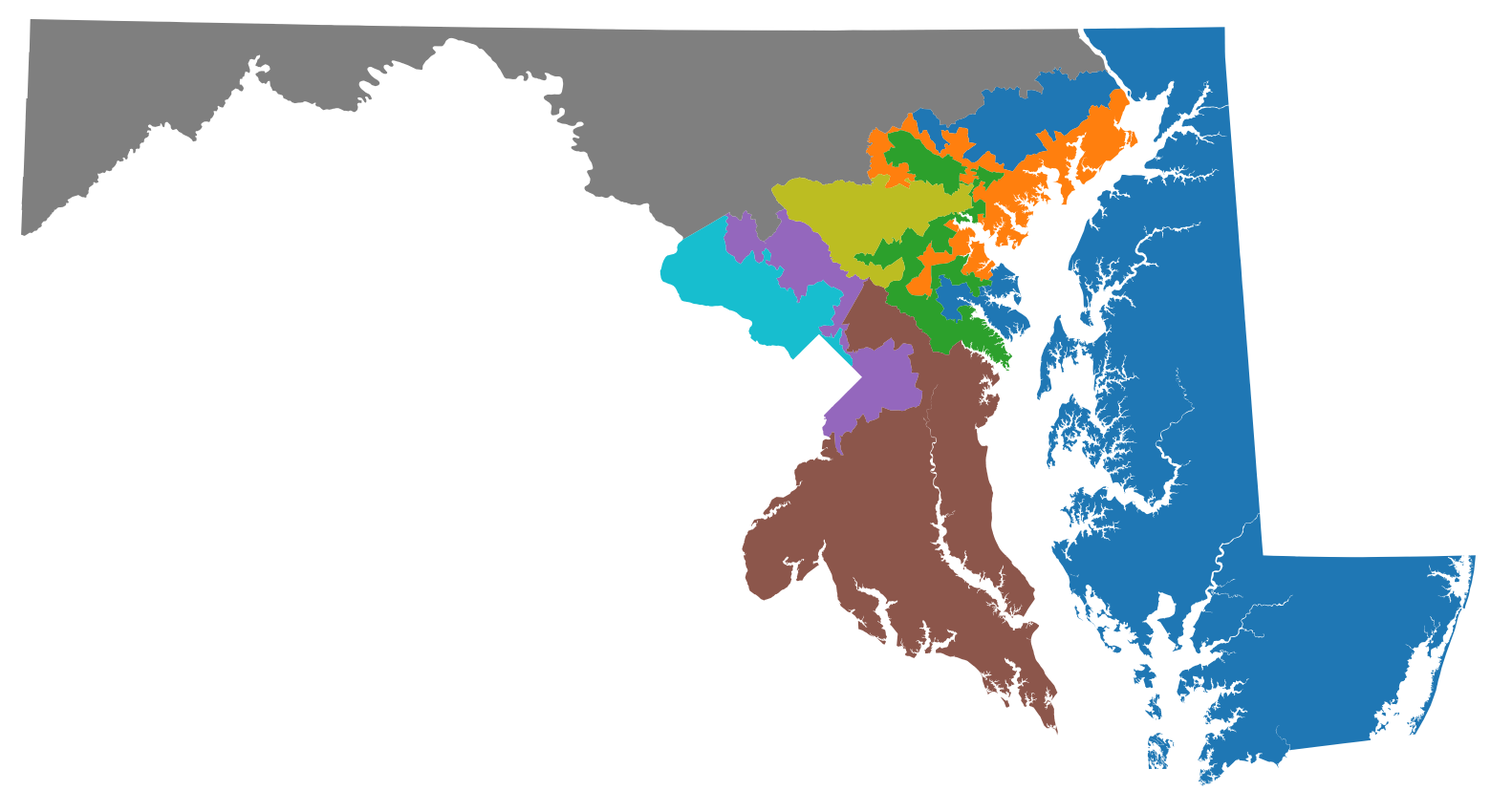}};
\node[inner sep=0pt, below=of MD103, align=right] (MD113)
    {\includegraphics[width=0.4\textwidth]{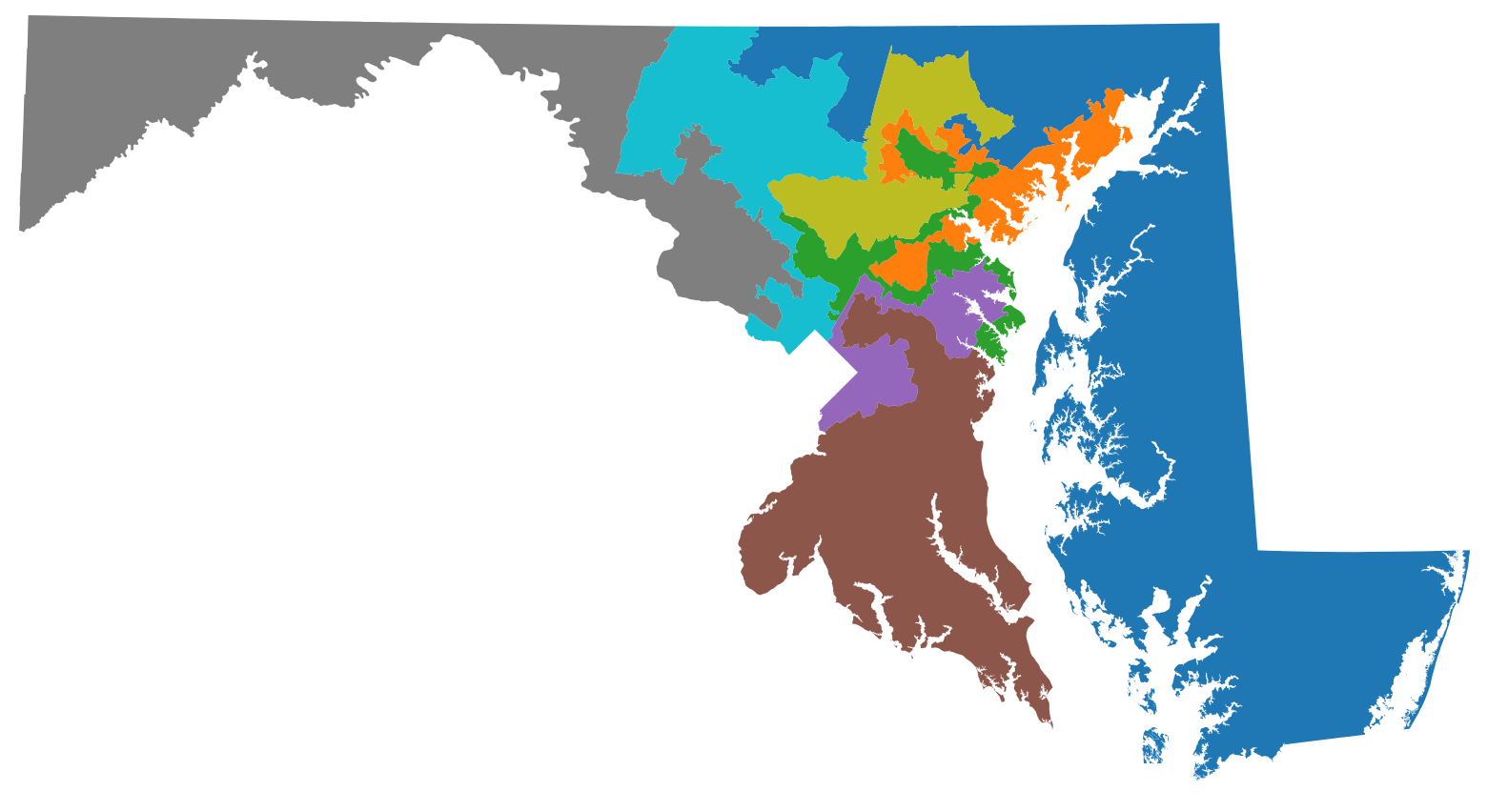}};
\end{scope}
\begin{scope}[node distance=0.2]
\node[below= of MD102] {1982-1991};
\node[below= of MD103] {1991-2002};
\node[below= of MD108]  {2002-2011};
\node[below= of MD113]  {2011-present};
\end{scope}
\draw[->,thick] (MD102.east) to node[above] {$\mathsf{Ent}=1.035$} (MD103.west);
\draw[->,thick] (MD103.south west) to node[above, sloped]  {$\mathsf{Ent}=0.957$} (MD108.north east);
\draw[->,thick] (MD108.east) to  node[above]  {$\mathsf{Ent}=1.117$} (MD113.west);
\end{tikzpicture}
\caption{MD congressional districts over three cycles with entropy distance}\label{fig:MDmaps}
\end{figure}
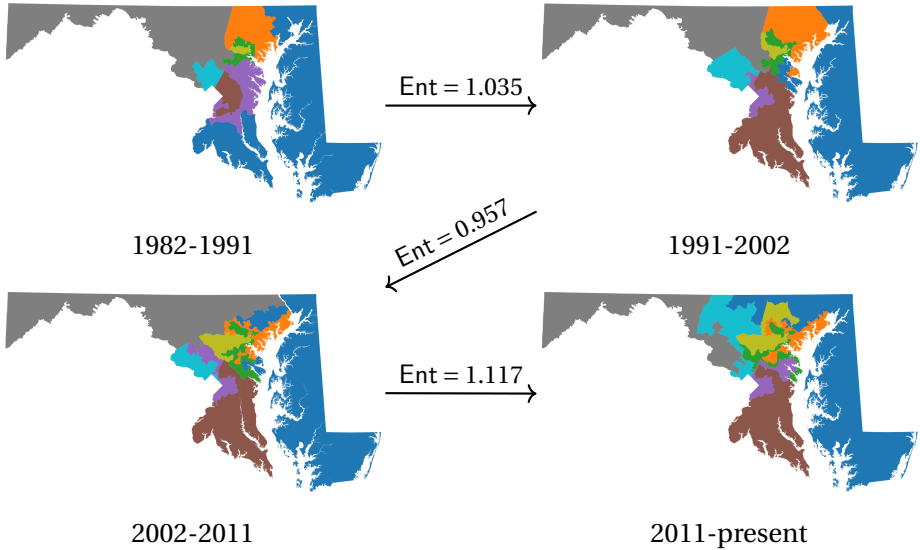

To illustrate the usefulness of conditional entropy in comparing plans, we consider three different redistrictings for the state of Maryland. \footnote{Census data obtained from NHGIS \cite{NHGIS}. Congressional districts obtained from \cite{UCLA} and matched to census blocks using the maup package (\url{github.com/mggg/maup}).} Figure \ref{fig:MDmaps} shows the four relevant districting plans, as well as the conditional entropy between the old and new plans for each time districts were redrawn. 

The entropy calculations seem to indicate that the 2011 redistricting brought the greatest amount of change to the congressional districts, a fact which is also visible on the maps to some extent. (Note also that this redistricting was the subject of a Supreme Court case.) The relationship between the 1991 and 2002 redistrictings is less clear just by looking at the maps. Remember that the entropy score weights regions strictly by population, not land area, so that smaller geographic details around large cities have a large influence on the entropy score, and these are hard to see with the naked eye. 

Returning to the eight Pennsylvania plans from the previous section, we can also compute the entropy distance of each plan to the 2011 plan. Table \ref{tab:PAdis} shows the results. The result for TS, GOV and REM are particularly interesting, since these three plans were all designed to replace the 2011 plan. Among these three, the TS plan has the lowest distance -- that is, it introduces the least change in districts (as measured in this particular way). In fact, the map-drawers of the TS plan made special mention of the fact that ``[the TS map] retains 68.8\% of the populations of existing districts in the same districts, which will help to reduce overall voter confusion'' \cite{TSstatement}.

\begin{table}
\centering
\begin{tabular}{|l|c|}
\hline
Plan $P$ & $\re{P}{2011}$  \\
\hline \hline
2011 & 0 \\
\hline
TS & 1.144 \\
\hline
GOV & 1.340\\
\hline
REM & 1.320 \\
\hline
8TH & 1.541\\
\hline
538 CPCT & 1.247 \\
\hline
538 GOP & 1.314 \\
\hline
538 DEM & 1.668 \\
\hline
\end{tabular}
\caption{Comparing PA plans to the 2011 enacted plan}
\label{tab:PAdis}
\end{table}

The difficulty of visually determining the extent of the difference between maps motivates the use of a precise mathematical score like conditional entropy. It is important to remember, however, that conditional entropy is insensitive to strictly geographic dissimilarity between plans because it is based only on population overlaps. It would fall to stakeholders in a given scenario to decide whether geographic or population-based similarity is more important. Even guidelines such as Nebraska's requirement that new districts ``preserve the cores of prior districts'' \cite{Nebraska} do not make the choice clear.

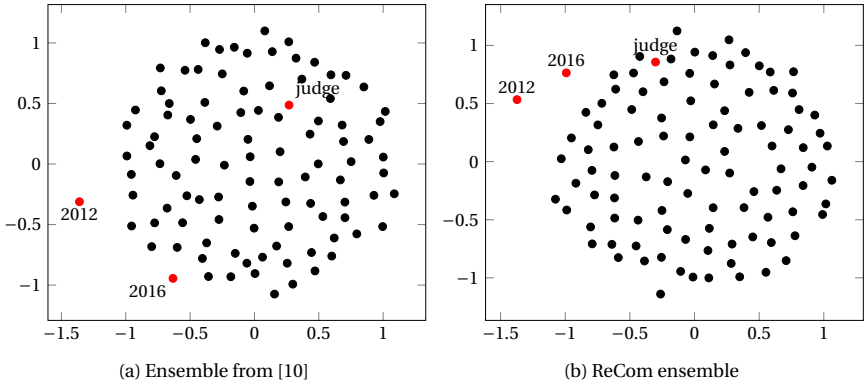
\begin{figure}
\centering
\begin{subfigure}{0.45 \textwidth}
\centering
\resizebox{\columnwidth}{!}{%
\begin{tikzpicture}
\begin{axis}[scatter/classes={a={red},b={black}}]
\addplot[scatter, only marks, scatter src=explicit symbolic] table[meta=label, x=X, y=Y, col sep=comma]{GuthNiehWeighill/MattinglyXY.csv}
node[pos=0, above right, text=black]{\small judge}
node[pos=0.01, below, text=black]{\small 2012}
node[pos=0.02, below left, text=black]{\small 2016};
\end{axis}
\end{tikzpicture}
}
\caption{Ensemble from \cite{Mattingly}}
\label{fig:mattscatter}
\end{subfigure}
\begin{subfigure}{0.45 \textwidth}
\centering
\resizebox{\columnwidth}{!}{%
\begin{tikzpicture}
\begin{axis}[scatter/classes={a={red},b={black}}]
\addplot[scatter, only marks, scatter src=explicit symbolic] table[meta=label, x=X, y=Y, col sep=comma]{GuthNiehWeighill/recomXY.csv}
node[pos=0, above, text=black]{\small judge}
node[pos=0.01, above, text=black]{\small 2012}
node[pos=0.02, above, text=black]{\small 2016};
\end{axis}
\end{tikzpicture}
}
\caption{ReCom ensemble}
\label{fig:recomscatter}
\end{subfigure}
\caption{Human (red) and computer-generated districting plans for North Carolina }
\label{NCplans}
\end{figure}

Another possible application of entropy as a distance between plans is to try to detect outliers. To demonstrate this, we consider the analysis of congressional redistricting in North Carolina performed by Mattingly et al. in \cite{Mattingly}. Using a large ensemble of computer generated plans, the authors demonstrate that the districting plans enacted in 2012 and 2016 were highly atypical outliers in terms of their partisan statistics (e.g. seats won for either party), while a third plan proposed by a panel of judges represents the ensemble of plans much more closely. Figure \ref{NCplans} shows the three human-made plans (2012, 2016 and the judges' plan) as well as one hundred plans drawn from the authors' ensemble using conditional entropy as the distance. The embedding is done using multi-dimensional scaling (MDS). We also repeat the analysis with an ensemble generated using a recombination Markov chain \cite{ReComPaper} \footnote{\url{github.com/mggg/gerrychain}}. For more details on Markov chains and ensemble methods, see DeFord and Duchin's chapter in \cite{book}.

Figure \ref{NCplans} seems to indicate that in addition to being outliers in \emph{partisan statistics}, the 2012 and 2016 plans are atypical (compared to the ensembles) with respect to how they divide up the population, while the judges' plan is more representative in this sense. Note that no vote data is involved in this kind of analysis since it depends only on the populations of the geographic units used, so that the outlier status of the 2012 and 2016 plans are independent of how the votes fall in any given election. This may or may not be an advantage of this kind of analysis, since a gerrymander which is geographically similar to the ensemble but manages to produce extreme partisan outcomes would not be flagged as an outlier.

Other distances between plans which can be found in the literature include a distance based on ideas from optimal transport \cite{transportpaper}. This distance is sensitive not just to the size of the overlap between pairs of districts (which completely determines the entropy score), but also to the geographic distance between pairs of districts, making it a far more sensitive measure of dissimilarity. The cost is that it takes longer to compute since an optimization problem must be solved to calculate the distance between two plans. A restriction to this distance (and others that require a perfect matching between the two plans) is that it is only defined for two plans with the same number of districts. Conditional entropy, on the other hand, does not require this condition and thus can be used in situations where a state gains or loses a congressional seat.

\section{Conclusion}
When Shannon developed his theory of communication at Bell Laboratories, he probably wasn't thinking about redistricting. Then again, neither was Markov, Bayes or any of the other great mathematicians and scientists whose work forms the basis of the modern study of redistricting, some of which is outlined in \cite{book}. While no mathematical concept is likely to be the perfect analytical framework for every possible situation -- this includes entropy, as we have remarked at various points in this chapter -- they can often provide valuable tools for analyzing the complex issues that arise every day in a democracy.

\section*{Appendix: Formal Definitions} \label{formal}

Our previous discussion has relied on the idea of an average number of bits of information, but we can also give a definition which is more straightforward to calculate. 

First, we define entropy for a single partition $X$. Suppose that $X$ divides the region into pieces of sizes $p_i$, normalized such that $\sum_i p_i = 1$. That is, $p_i$ is the fraction of voters in part $i$. Then the entropy of $X$ is
$$Ent{(X)} = \sum_i p_i \log_2 \left( \frac{1}{p_i} \right).$$
Intuitively, a higher value of $Ent{(X)}$ means that the partition divides the region into smaller pieces. As mentioned above, we can also think of $Ent{(X)}$ as the average number of bits needed to communicate which part of $X$ a voter is in.

What about conditional entropy? For two partitions $X$ and $Y$, the conditional entropy Ent$({X}|{Y})$ is a weighted sum of entropies calculated over each part of $Y$. This definition is a bit technical, but the underlying concept is simple: just take the entropy of each piece, and add them up with weights proportional to the population.

For each part $Y_j$ of $Y$, the partition $X$ also induces a partition of $Y_j$. The parts of this induced partition are $X_i \cap Y_j$ for each part $X_i$ of $X$. (For example, if $X$ is a districting plan, and $Y$ is the county partition, the induced partition describes how the districting plan partitions county $j$.) 

We define Ent${(X|Y_j)}$ to be the entropy of this induced partition. That is, $$ Ent{(X | Y_j)} = \sum_i p_{ij} \log_2 \left( \frac{1}{p_{ij}} \right)$$ where $p_{ij}$ is the fraction of the population of $Y_j$ that lies in $X_i$. You can think of Ent${(X|Y_j)}$ as sort of a ``local entropy'' of the plan $X$ when restricted to only $Y_j$.

Now, to calculate the conditional entropy, we simply add up the local entropies weighted by population: $$Ent({X}|{Y}) = \sum_j q_j  \, \ent{X | Y_j}$$ where the coefficients $q_j$ are the sizes of the parts $Y_j$. 

For a concrete example, consider the districts and counties application we had earlier. To calculate Ent${(D}|{C)}$ for districts $D$ and counties $C$, we can use the formula

$$
Ent{(D}|{C)} = \sum_{c \in \cou} \frac{\pop(c)}{T} \sum_{d \in \dis} \frac{\pop(c \cap d)}{\pop(c)} \log_2 \left ( \frac{\pop (c)}{\pop(c \cap d)} \right )
$$
where $T$ denotes the total population of the state.

Finally, a note on Ent${(X}|{Y)}$ versus Ent${(Y}|{X)}$. These two quantities differ in general: the former asks how well $Y$ predicts $X$ while the latter asks how well $X$ predicts $Y$. In the case of counties and districts, for example, Ent${(D}|{C)}$ will be zero if every county is contained in only one district, but Ent${(C}|{D)}$ will typically be non-zero since each district may have more than one county contained in it. Nonetheless, there is the following relationship:

\begin{align*}
 \mathit{(Bayes' Rule}) \quad \re{X}{Y} + \ent{Y} = \re{Y}{X} + \ent{X}
\end{align*}

Intuitively, this holds because both sides of the equation are the number of bits of information required to encode both $X$ \emph{and} $Y$. 

\section*{Acknowledgements}
The authors would like to thank Colby Brown and Brandon Kolstoe, two students at the Voting Rights Data Institute 2019, for their help with the experiments in this chapter. We would also like to thank Ruth Buck for help with the geographic data required for the experiments. We would like to thank Olivia Walch for the illustrations in this chapter. Thomas Weighill acknowledges the support of the NSF Convergence Accelerator Grant No. OIA-1937095.